\documentclass[prl,twocolumn]{revtex4}
\usepackage[dvips]{graphics,color}
\usepackage{amsmath}
\usepackage{amsfonts}

\newcommand{\tr}{\mbox{tr}}
\usepackage{amssymb}
\usepackage{amstext}
\usepackage[sort&compress]{natbib}
\newcommand{\ignore}[1]{}

\newcommand{\ket}[1]{| #1 \rangle}
\newcommand{\bra}[1]{\langle #1 |}
\begin{document}

\title{Exploiting non-quantum entanglement to widen applicability of limited-entanglement classical simulations of quantum systems.}

\author{N. Ratanje and S. Virmani}

\affiliation{Department of Physics SUPA, University of Strathclyde, John Anderson Building,
107 Rottenrow, Glasgow, G4 0NG, United Kingdom}

\date{\today}

\begin{abstract}
It is known that if the quantum gates in a proposed quantum computer are so noisy that they are incapable of generating
entanglement, then the device can be efficiently simulated classically. If the measurements and single particle
operations are restricted, then the same statement can be true for generalised non-quantum notions of entanglement.
Here we show that this can improve the applicability of limited-entanglement simulation algorithms. In particular, we show that a classical simulation algorithm of Harrow \& Nielsen can efficiently simulate magic state quantum computers that are ideal other than noise on the CNOTs for joint depolarising strengths of $272/489 \sim 56\%$, in contrast to noise levels of $2/3 \sim 66\%$ required if the algorithm uses quantum notions of separability. This suggests that quantum entanglement may not be the most appropriate notion of entanglement to use when discussing the power of stabilizer based quantum computers.
\end{abstract}

\maketitle

\section{Introduction}

Entanglement is often cited as the main reason why quantum devices cannot be efficiently simulated on classical computers. However, while it is the case that devices with limited entanglement can often be efficiently classically simulated \cite{liment,HN,ABIN,VHP}, there can also be highly entangled systems that can be efficiently simulated classically \cite{GK,ENT}. The most prominent example of this arises in fault tolerant quantum computation. In such architectures the stabilizer components generate the quantum entanglement, but acting alone they can be efficiently simulated classically \cite{GK}. Universal quantum computation is usually achieved by injecting a single qubit state, gate, or measurement into the circuit. So fault tolerant quantum computation appears to provide a regime where entanglement is not the source of the computational power.

Motivated by this example, in this paper we will present arguments that ordinary {\it quantum} entanglement may not be the most appropriate notion of entanglement to use when discussing the computational power of stabilizer based computations. We will do this by showing that using a {\it generalised} notion of entanglement, as discussed in \cite{Barnum}, we may significantly enhance the range of applicability of classical simulation algorithms that rely on limited entanglement. In particular we will consider applying the classical simulation algorithm of Harrow \& Nielsen \cite{HN} to quantum computers built using stabilizer components and non-stabilizer single qubit resources (the so-called `magic-state' architectures \cite{BK}). If quantum entanglement is used, the Harrow-Nielsen algorithm requires that the CNOT undergo $2/3\sim$66\% of joint depolarising noise in order to provide an efficient classical simulation. However, we will show that in the case of magic-state architectures, an alternative notion of entanglement can improve the range of applicability of the algorithm to less than $56$\% noise on the CNOT. In the same spirit as the research programme proposed by \cite{Barnum}, this suggests that in the context of fault tolerant quantum computation there may be different notions of entanglement that are more appropriate when discussing computational power.

Note that our work is distinct from the usual approach to analysing noise induced classicality in magic-state computers,
which treats the non-stabiliser resource as the source of non-classicality, and aims to to understand the noise levels required
before this resource loses its power \cite{BK,magicmore}. Here we are treating the single qubit non-stabilizer resources as classical, but assessing the power
brought by the generalised-entangling power of the CNOT.

\section{Generalised entanglement and classical simulation}

A quantum state of two particles $A$ and $B$ is said to be quantum separable if it can be written as a probabilistic mixture of
products of single particle quantum states $\rho_A \otimes \rho_B$. In the context of qubits, $\rho_A$ and $\rho_B$
must be quantum states drawn from the Bloch sphere, which we denote by the symbol $Q$. This conventional definition of
entanglement can be generalised \cite{Barnum}: for any convex set $C$ of single particle Bloch vectors (perhaps including ones outside the Bloch sphere), we define an operator to be $C$-separable if it can be decomposed as a probabilistic mixture of product `states' $\rho_A \otimes \rho_B$ such that $\rho_A,\rho_B \in C$.

The notion of quantum separability is utilised in the algorithm developed by Harrow \& Nielsen \cite{HN} to classically simulate devices with non-entangling gates. The algorithm samples the separable decomposition of the system after each gate has been applied, and stores the product state that results from each such sampling step. The final result of the computation is then obtained by sampling the probabilities of measurement outcomes on each individual particle. The algorithm is efficient as it only ever stores and manipulates product operators, which is computationally inexpensive.

If we allow all single particle measurements, then to be able to perform the final sampling step it is essential that the variables stored correspond to single particle quantum states - otherwise the final sampling will involve negative probability distributions. However, if the measurements in the device are {\it restricted}, then the HN algorithm will work even if the gates are non-entangling w.r.t. any set $C$ that is contained within the normalised dual $M^*$ of the set of available measurement operators $M$ (see e.g. \cite{RV}):
\begin{equation}
C \subset M^* := \{\rho| \tr\{\rho\}=1, \tr\{\rho B\} \geq 0, \forall B \in M \},
\end{equation}
In the case where $M$ is the set of all quantum measurements, then $M^*$ is the set of quantum states. If $M$ is
a restricted, then $M^*$ will be correspondingly larger. The notion of entanglement changes accordingly \cite{Barnum,Boxworld}. For example,
if $M$ is the set of Pauli measurements, then Bell pairs become separable (see e.g. \cite{RV}).

[Aside: It is important to note one subtlety: for the classical simulation to function with such a modified notion of entanglement we also require that the {\it single particle} operations cannot produce single particle
states outside $C$, i.e. that the single particle states can only be initialised within $C$, and $C$ must remain invariant
under the action of any single particle gates.]

In \cite{RV} such considerations were applied to understand the effects of noise on computation in (among other situations) magic state quantum computation, in which single particle unitaries are drawn from the Clifford group and the measurements $M$ are in Pauli directions only. While this did provide new regimes that could be efficiently simulated classically, it did not lead to significant improvements in the applicability of the HN algorithm, in that the levels of noise required to remove the entangling power of the CNOT were typically no lower than those required to remove quantum entanglement.

However in \cite{RV} the set $C$ was taken to be the {\it full} dual, i.e. $C=M^*$, and this is not required as we can only ever initialise the qubits in genuine quantum states. So in fact one can consider choices of $C$ such that $Q \subset C \subset M^*$. In this work we show that in the context of magic state architectures there exist choices of $C$ such that $Q \subset C \subset M^*$ that can lead to quite large differences in the entangling power of the CNOT (CX) or CSIGN (CZ) gate. In particular we find that the CZ gate requires only 56\% depolarising noise to remove its entangling power, in contrast to 2/3 = 66\% if we choose $C=Q$.

\section{Notation}

We will represent two qubit operators in one of two ways, both relying upon the Pauli expansion:
\begin{equation}
{1 \over 4} \sum_{i,j=0,..,3} \rho_{ij} \sigma_{i} \otimes \sigma_{j}
\end{equation}
where $\sigma_{0},\sigma_{1},\sigma_{2},\sigma_{3}$ represent the 2x2 Identity
and the Pauli X,Y,Z operators respectively. In fact as we will only consider trace
normalised operators, we will always have $\rho_{00}=1$.
We will display the
coefficients $\{\rho_{ij}\}$ as a 4 x 4 matrix, with rows/columns numbered
from 0,..,3:
\begin{eqnarray}\left(\begin{array}{cccc}
     \rho_{00}=1 & \rho_{01} & \rho_{02} & \rho_{03} \\
     \rho_{10} & \rho_{11} & \rho_{12} & \rho_{13} \\
     \rho_{20} & \rho_{21} & \rho_{22} & \rho_{23} \\
     \rho_{30} & \rho_{31} & \rho_{32} & \rho_{33}
      \end{array}\right) \label{matrix}
\end{eqnarray}
or as a column vector formed by sequentially concatenating the columns of this matrix, i.e.
\begin{equation}
(\rho_{00},\rho_{10},\rho_{20},\rho_{30},\rho_{01},..,\rho_{31},\rho_{02},..,\rho_{32},\rho_{03},..,\rho_{33})^T. \label{column}
\end{equation}
Please note that we do not represent 2-qubit operators as density matrices
in the computational basis - we always use this Pauli expansion form.

Although the results can be modified straightforwardly to any 2-qubit
Clifford gate, the only gate that we will consider in this work is the CSIGN or CZ gate, which is
the unitary $\ket{0}\bra{0}\otimes I + \ket{1}\bra{1}\otimes Z$.

We will use the symbol ${\cal C}_\lambda$
to denote an ideal CZ gate followed by joint depolarising noise, where with probability $\lambda$ the output of the gate is replaced with a maximally mixed state represented by a normalized identity.

\begin{figure}[t]
        \resizebox{7cm}{!}{\includegraphics{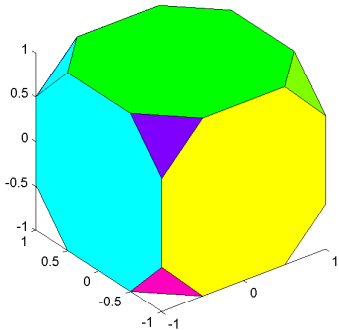}}
        \caption{A truncated cube.}
        \label{shape}
\end{figure}

The sets $C$ that we consider consider here will be truncated cubes of Bloch vectors. We have investigated
a number of sets that are invariant under the single-qubit Clifford group, and truncated cubes are the best of these
in the parameter ranges of interest. The truncated cubes (see fig. (\ref{shape})) are parameterised
by a variable $r \in (0,1]$ and are defined in the following way: $TRUN(r)$ is the truncated cube given by the convex hull of Bloch vectors $(x,y,z)$ of the form
$(|x|,|y|,|z|)=(1,1,r),(1,r,1),(r,1,1)$, i.e. the convex hull of all possible sign choices and choices of either $|x|,|y|$ or $|z|$ equal to $r$. There are hence 24 extremal points of $TRUN(r)$. For two particles $A$ and $B$ will need to consider products of
states selected from $TRUN(r)$.  We will use the shorthand $(x_A,y_A,z_A)\otimes(x_B,y_B,z_B)$ to denote to such states,
where $(x_A,y_A,z_A)$ and $(x_B,y_B,z_B)$ are the individual Bloch vectors. In particular the maximally
mixed state can be represented as $(0,0,0)\otimes(0,0,0)$.

\section{Problem and basic method}

Our goal is to calculate, for various choices of $C=TRUN(r)$, the level of joint depolarising noise
required to make all output states separable w.r.t. $TRUN(r)$. In particular we must go through all
possible pairs of input extremal points, act upon them with ${\cal C}_\lambda$ and find the minimal $\lambda$
required to make sure that all outputs are separable. There are 24$\times$24 possible pairs of extremal input states,
but in fact it is sufficient to consider only four of them. The reason for this is that if a given operator
$\rho$ is separable w.r.t. $TRUN(r)$, then so is any operator obtained from its matrix
representation (\ref{matrix}) by (a) permuting
any of the last three columns, (b) permuting any of the last three rows, (c) transposition of the matrix, (d)
multiplying any of the last three rows or columns by minus signs. This is because all these operations
are linear, and leave $TRUN(r)$ invariant. Up to these transformations the outputs obtained from the full set of pairs of extremal input states are equivalent to the outputs of only four pairs, which we refer to as Cases 1 to 4:
\begin{eqnarray}
\rho_1:= (1,1,r) \otimes (1,1,r) \,\,
\rho_2:= (r,1,1) \otimes (1,1,r)  \nonumber\\
\rho_3:= (1,1,r) \otimes (1,r,1) \,\,
\rho_4:= (r,1,1) \otimes (1,r,1) \nonumber
\end{eqnarray}
Hence our goal is to compute the minimal $\lambda$ required to make sure that the outputs ${\cal C}_\lambda(\rho_i)$ are $TRUN(r)$ separable
for $i=1,..,4$. Consider for instance Case 1. We must compute:
\begin{eqnarray}
\min && \lambda \nonumber \\
{\rm{s.t.}}\,\,\,&& (1-\lambda) CZ(\rho_1) + \lambda ((0,0,0) \otimes (0,0,0)) \nonumber \\
&& \in TRUN(r)-sep \nonumber
\end{eqnarray}
This is a linear programming problem, and hence
can be readily solved using standard numerical tools. In fact, as we shall see it turns out that the structure of
the problem enables us to use computational algebra packages to obtain almost completely
analytic solutions. We can reexpress the linear programme as:
\begin{eqnarray}
\min && p_0 \nonumber \\
{\rm{s.t.}}\,\,\,
&& CZ(\rho_1) = \sum_{n=0,..,576} p_n E_n , \label{sepcond} \\
&& p_n \geq 0 \label{probcond}
\end{eqnarray}
where
\begin{eqnarray}
E_0 := (CZ(\rho_1) - (0,0,0) \otimes (0,0,0)) \,\,\, , \,\,\,p_0 := \lambda \nonumber
\end{eqnarray}
and for $n=1,..,576$ the $E_n$ denote pairs of extremal product states from $TRUN(r)$, and the $p_n$ represent
the probabilities with which they occur in the separable decomposition. We do not need to impose normalisation
$\sum_{n>0} p_n=1$, as this is implicitly required by the other conditions.

A numerical linear programming routine will usually return the minimal value of $p_0$ (i.e. $\lambda$) as well
as the values of $p_n$ (for $n>0$) that achieve it. For our purposes it is important to note
that the values of $p_n$ that achieve the optimum can always be chosen to
be non-zero only on a {\it linearly independent} subset of vectors from $E_n$.
The reason for this that the optimum can always be achieved on an extremal
point of the convex set of $p_n$ that satisfy the constraints,
and it is well known in linear programming theory that the extremal points must be non-zero only on linearly independent subsets of
$E_n$.

One can find such a linearly independent solution by first solving the the programme
numerically, and then one by one removing the vectors $E_n$, each time checking whether the optimal
solution can still be attained. If at any step the removal of a particular $E_n$ prevents us
from reaching the optimal solution, we put that vector back into the problem and carry on removing
others. Eventually we terminate at an optimal solution involving only a linearly independent subset
of $E_n$.

This numerical solution can then be turned into an analytic proof that the value of $\lambda$ is
{\it achievable}. One simply takes the symbolic version of the matrix formed from the optimal linearly
independent subset of $E_n$, and then inverts the condition (\ref{sepcond})
to obtain an analytic expression for the optimal $p_n$. The solutions obtained in this way are
not usually simple, as the optimal separable decompositions are usually not unique, and so the computer
does not necessarily identify the neatest solution. However, for completeness in the appendix we present example
decompositions for the region around $r=1/2$.

While this demonstrates achievability analytically, it does not demonstrate necessity analytically - i.e. that
the solution is optimal. One approach to attempt to fill this gap is to obtain a set of inequalities
(the analogue of entanglement witnesses) that are necessary conditions for any operator to be $TRUN(r)$-separable,
and then turn these into inequalities on $\lambda$ that could match the achievable values of $\lambda$.
 In the next section we will see that in the regime around $r=1/2$ (which appears to be the best region) we
have been able to find suitable inequalities for Cases 2-4 above, and hence the optimal $\lambda$ for these cases analytically. In case
1 we have not been able to identify suitable inequalities. However, putting all four cases together for $r=1/2$ we find
that $\lambda=272/489 \sim 0.56$ is analytically achievable, whereas $\lambda=5/9 \sim 0.55$ is necessary. So our analytic solutions
are essentially tight.

\section{Results and Necessary conditions}

\begin{figure}[t]
        \resizebox{13cm}{!}{\includegraphics{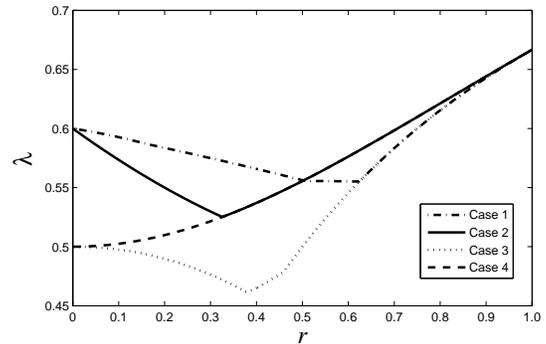}}
        \caption{Plot of the minimal noise level $\lambda$ required to make the output of a noisy CZ $TRUN(r)$ separable,
        for each of the four cases. For all values of $r$, Case 4 matches its lower bound in equation (\ref{4lower}).
        For values of $r$ decreasing from $r=1$ the other three cases also match the lower bounds (\ref{3lower},\ref{2lower}),
        but for each of these cases there is a value of $r$ below which these bounds become unattainable.}
        \label{lambdaplot}
\end{figure}

As the truncated cubes that we consider are within the dual of ideal Pauli measurements, it is a necessary condition
for separability that the output operators return positive values for the probabilities of measurement
outcomes for the various Pauli operators. It turns out that for values of $r$ or most interest such inequalities
are For instance, consider Case 4. Under the action of ${\cal C}_\lambda$ the output operator is:
\begin{eqnarray}\left(\begin{array}{cccc}
     1 &  (1-\lambda) r & 1-\lambda & 1-\lambda \\
    1-\lambda & (1-\lambda) r & -(1-\lambda)r^2 & 1-\lambda\\
      (1-\lambda) r &  -(1-\lambda)  &  (1-\lambda) r &  (1-\lambda) r \\
     1-\lambda &  (1-\lambda) r & 1-\lambda & 1-\lambda
      \end{array}\right)
\end{eqnarray}
On this operator the requirement of having a positive probability of getting down, down in a measurement of
the $X$ and $Y$ operators leads to:
\begin{equation}
\lambda \geq 1- (1 /(2+r^2)). \label{4lower}
\end{equation}
Performing identical computations for Cases 1-3 gives:
\begin{eqnarray}
{\rm{Cases 1 \& 3}}: \,\,\,\lambda \geq 1- (1 / (1+2r))  \label{3lower} \\
{\rm{Case 2}}: \,\,\,\lambda \geq 1- (1 / (2+r^2)) \label{2lower}
\end{eqnarray}
We will now see that in cases 2-4 these inequalities are tight. The numerical results for
Cases 1-4 are plotted in figure \ref{lambdaplot}. For all values of $r$, Case 4 matches its lower bound in equation (\ref{4lower}). For values of $r$ decreasing from $r=1$ the other three cases also match the lower bounds (\ref{3lower},\ref{2lower}), but for each of these cases there is a value of $r$ below which these bounds become unattainable.

We see from these results that $r=1/2$ appears to require the least noise. This value of $r$ is also large
enough to include the whole Bloch sphere, and hence all magic-states. At the point $r=1/2$, Cases 2,3 and 4
attain the lower bounds (\ref{2lower}),(\ref{3lower}),(\ref{4lower}), hence showing that the higher of these values $
1-1/(2+r^2)= 5/9 = 0.5555$
is necessary and sufficient for these cases. With Case 1 we have not been able to identify matching lower bounds,
however the method of the previous section can be used to show that in the vicinity of $r=1/2$ we may achieve values of
\begin{equation}
\lambda = (4\, r^2 + 8\, r + 12)/({r^4 + 2\, r^3 + 9\, r^2 + 16\, r + 20}). \nonumber
\end{equation}
At $r=1/2$ this evaluates to $272/489=0.5562$.

\section{Summary and Open problems}

We have shown that using a non-quantum notion of entanglement can increase the range of application
of Harrow \& Nielsen's limited entanglement simulation method. However, we do not know whether this persists for more general noise
models, or in the context of multiparticle entanglement.

We also do not know whether the truncated cubes are the optimal set for our problem. In future work \cite{RVprep} we will present lower bounds on the $\lambda$ required for any set $C$, where for any $Q \subset C \subset M^*$ we can show that $\lambda$ must be greater than 0.42. However, truncated
cubes are currently the best among the choices of $C$ that we have tried. In \cite{RVprep} we will also present an analysis of a generalised problem where the sets $Q$ and $M^*$ are reduced and expanded respectively
by noise. In very noisy regimes one can identify the optimal sets that require the least amount of depolarising noise to make the CZ separable (they turn out to be either spheres or octahedra), but for
the most interesting case of noise-free $Q$ and $M^*$ it open whether truncated cubes are the best choice.

\section{Acknowledgments}

This work is supported by EU STREP `Corner', a University of Strathclyde Starter grant, and an EPSRC PhD studentship.
We thank Jon Barrett, Steve Barnett, Vedran Dunjko, and Mark Howard for discussions.

\newpage

\begin{widetext}

\section{Appendix: Computer generated expressions for achievable values of $\lambda$ near $r=1/2$.}

In this appendix, for $r=1/2$, we present $TRUN(r)$ separable decompositions that show that the numerically obtained values
of $\lambda$ are achievable. The decompositions have been obtained using the computational methods described in the main text.
They are extremely messy objects - this may be a reflection of the fact that the optimal separable decompositions are usually
non-unique. It is possible that other methods may lead to more concise and illustrative expressions. Indeed, the solution
for Case 4 was guessed by hand, and the expressions are much simpler than the remaining cases.

\section{Case 1}
For Case 1, and a noise level given by
\begin{equation}
\lambda = 1 - \frac{r^4 + 2\, r^3 + 5\, r^2 + 8\, r + 8}{r^4 + 2\, r^3 + 9\, r^2 + 16\, r + 20}
\end{equation}
the output state of ${\cal C}_\lambda$ is given by (using the column vector notation of equation (\ref{column})):
 \begin{eqnarray}
\frac{r^4 + 2\, r^3 + 5\, r^2 + 8\, r + 8}{r^4 + 2\, r^3 + 9\, r^2 + 16\, r + 20}  \left(\begin{array}{c} \frac{r^4 + 2\, r^3 + 9\, r^2 + 16\, r + 20}{r^4 + 2\, r^3 + 5\, r^2 + 8\, r + 8} \\
r
\\ r
\\ r
\\ r
\\ 1
\\ -1
\\ 1
\\ r
\\ -1
\\ 1
\\ 1
\\ r
\\ 1
\\ 1
\\ r^2 \end{array}\right)
\end{eqnarray}
The separable decomposition of this state is given by the following expression, where the columns of the left matrix correspond
to product operator extrema (in the notation of equation (\ref{column})), and the elements of the right column vector
are the associated probabilities. In the vicinity of $r=1/2$ all the probabilities are positive, and so the decomposition
works for a range of $r$.
\begin{eqnarray}
\left(\begin{array}{ccccccccccccccc} 1 & -1 & 1 & 1 & 1 & -1 & 1 & -1 & 1 & 1 & -1 & -1 & r & - r & - r\\ r & - r & r & r & - r & - r & 1 & 1 & -1 & -1 & 1 & 1 & 1 & -1 & -1\\ -1 & -1 & 1 & 1 & 1 & 1 & - r & - r & r & - r & r & - r & 1 & -1 & 1\\ - r & - r & 1 & -1 & -1 & 1 & -1 & 1 & 1 & 1 & -1 & -1 & r & - r & 1\\ - r & r & 1 & -1 & -1 & -1 & -1 & -1 & 1 & 1 & 1 & 1 & r^2 & r^2 & - r\\ - r^2 & r^2 & r & - r & r & - r & -1 & 1 & -1 & -1 & -1 & -1 & r & r & -1\\ r & r & 1 & -1 & -1 & 1 & r & - r & r & - r & - r & r & r & r & 1\\ -1 & -1 & r & r & r & 1 & - r & - r & -1 & -1 & 1 & 1 & 1 & -1 & r\\ -1 & 1 & r & r & r & -1 & - r & r & -1 & -1 & -1 & -1 & r & r & - r^2\\ - r & r & r^2 & r^2 & - r^2 & - r & - r & - r & 1 & 1 & 1 & 1 & 1 & 1 & - r\\ 1 & 1 & r & r & r & 1 & r^2 & r^2 & - r & r & r & - r & 1 & 1 & r\\ 1 & -1 & 1 & 1 & -1 & - r & 1 & 1 & r & - r & r & - r & 1 & -1 & -1\\ 1 & 1 & 1 & 1 & -1 & r & 1 & -1 & r & - r & - r & r & r & r & r\\ r & r & r & r & r & r^2 & 1 & 1 & - r & r & r & - r & 1 & 1 & 1\\ -1 & 1 & 1 & 1 & -1 & - r & - r & - r & r^2 & r^2 & r^2 & r^2 & 1 & 1 & -1 \end{array}\right)
\left(\begin{array}{c} \frac{2\, \left(r^2 - 2\, r + 4\right)}{ - r^5 - 5\, r^3 + 2\, r^2 + 12\, r + 40}\\ 0\\ \frac{4\, \left(r^2 + r + 1\right)}{\left(r + 1\right)\, \left(r^4 + 2\, r^3 + 9\, r^2 + 16\, r + 20\right)}\\ 0\\ 0\\ -\frac{2\, r^2\, \left(r - 1\right)}{ - r^5 - 5\, r^3 + 2\, r^2 + 12\, r + 40}\\ -\frac{2\, r^2\, \left(r - 1\right)}{ - r^5 - 5\, r^3 + 2\, r^2 + 12\, r + 40}\\ 0\\ \frac{ - r^6 + r^5 + 12\, r^3 + 21\, r^2 + 27\, r - 12}{4\, \left(r^5 + 3\, r^4 + 11\, r^3 + 25\, r^2 + 36\, r + 20\right)}\\ -\frac{r^7 - r^6 + 2\, r^5 - 8\, r^4 - 9\, r^3 + r^2 + 54\, r + 8}{4\, \left(r^6 + r^5 + 5\, r^4 + 3\, r^3 - 14\, r^2 - 52\, r - 40\right)}\\ \frac{ - r^6 - 2\, r^4 + 6\, r^3 - r^2 + 14\, r + 8}{4\, \left( - r^5 - 5\, r^3 + 2\, r^2 + 12\, r + 40\right)}\\ -\frac{ - r^7 + r^6 - 6\, r^5 + 12\, r^4 - 11\, r^3 + 11\, r^2 + 2\, r + 40}{4\, \left(r^6 + r^5 + 5\, r^4 + 3\, r^3 - 14\, r^2 - 52\, r - 40\right)}\\ -\frac{r\, \left( - r^5 - 2\, r^3 + 6\, r^2 + 5\, r + 16\right)}{r^6 + r^5 + 5\, r^4 + 3\, r^3 - 14\, r^2 - 52\, r - 40}\\ \frac{ - r^6 - 2\, r^4 + 6\, r^3 + 5\, r^2 + 12\, r - 8}{\left(r + 1\right)\, \left(r - 2\right)\, \left(r^4 + 2\, r^3 + 9\, r^2 + 16\, r + 20\right)}\\ \frac{2\, \left(r^2 - 2\, r + 4\right)}{ - r^5 - 5\, r^3 + 2\, r^2 + 12\, r + 40} \end{array}\right) \nonumber
\end{eqnarray}

\section{Case 2}

For Case 2, in the vicinity of $r=1/2$, and for a noise level given by
\begin{equation}
\lambda = 1 - {1 \over 2 +r^2}
\end{equation}
the output state of ${\cal C}_\lambda$ is given by (using the column vector notation of equation (\ref{column})):
\begin{eqnarray}
{1 \over 2 +r^2} \left(\begin{array}{c} 2 +r^2 \\ r^2\\ r\\ 1\\ 1\\ 1\\ - r\\ 1\\ 1\\ -1\\ r\\ 1\\ r\\ r\\ 1\\ r \end{array}\right)
\end{eqnarray}
The probabilities $p_m$ appearing in the decomposition are given by:
\begin{eqnarray}
\left(\begin{array}{c} -\frac{3\, r^{17} + 9\, r^{16} - 44\, r^{15} - 121\, r^{14} - 92\, r^{13} - 49\, r^{12} + 692\, r^{11} + 807\, r^{10} + 1458\, r^9 + 305\, r^8 + 1020\, r^7 - 419\, r^6 + 772\, r^5 - 371\, r^4 + 252\, r^3 - 139\, r^2 + 35\, r - 22}{4\, r\, \left(r^2 + 2\right)\, \left(r^{15} + 5\, r^{14} + 2\, r^{13} + 12\, r^{12} + 47\, r^{11} - 129\, r^{10} - 88\, r^9 - 122\, r^8 + 15\, r^7 + 99\, r^6 + 26\, r^5 + 96\, r^4 + r^3 + 25\, r^2 - 4\, r + 14\right)}\\ \frac{r^{17} + 5\, r^{16} + 22\, r^{15} + 73\, r^{14} + 98\, r^{13} + 447\, r^{12} - 154\, r^{11} + 805\, r^{10} - 476\, r^9 + 1245\, r^8 - 766\, r^7 + 1131\, r^6 - 642\, r^5 + 357\, r^4 - 126\, r^3 + 39\, r^2 - 5\, r - 6}{4\, r\, \left(r^2 + 2\right)\, \left(r^{14} + 4\, r^{13} - 2\, r^{12} + 14\, r^{11} + 33\, r^{10} - 162\, r^9 + 74\, r^8 - 196\, r^7 + 211\, r^6 - 112\, r^5 + 138\, r^4 - 42\, r^3 + 43\, r^2 - 18\, r + 14\right)}\\ -\frac{ - r^{18} - 4\, r^{17} + 17\, r^{16} + 87\, r^{15} + 159\, r^{14} + 297\, r^{13} + 355\, r^{12} + 419\, r^{11} + 1041\, r^{10} + 405\, r^9 + 1201\, r^8 - 331\, r^7 + 917\, r^6 - 661\, r^5 + 353\, r^4 - 175\, r^3 + 60\, r^2 - 37\, r - 6}{4\, r\, \left(r^2 + 2\right)\, \left(r^{15} + 5\, r^{14} + 2\, r^{13} + 12\, r^{12} + 47\, r^{11} - 129\, r^{10} - 88\, r^9 - 122\, r^8 + 15\, r^7 + 99\, r^6 + 26\, r^5 + 96\, r^4 + r^3 + 25\, r^2 - 4\, r + 14\right)}\\ -\frac{ - 2\, r^{15} - r^{14} + 3\, r^{13} + 60\, r^{12} + 72\, r^{11} + 233\, r^{10} + 293\, r^9 + 406\, r^8 + 38\, r^7 + 289\, r^6 - 259\, r^5 + 48\, r^4 - 108\, r^3 - 9\, r^2 - 37\, r - 2}{2\, \left(r^2 + 2\right)\, \left( - r^{14} - 6\, r^{13} - 8\, r^{12} - 20\, r^{11} - 67\, r^{10} + 62\, r^9 + 150\, r^8 + 272\, r^7 + 257\, r^6 + 158\, r^5 + 132\, r^4 + 36\, r^3 + 35\, r^2 + 10\, r + 14\right)}\\ \frac{r^{17} + 11\, r^{16} + 35\, r^{15} + 55\, r^{14} + 125\, r^{13} + 17\, r^{12} + 175\, r^{11} + 219\, r^{10} + 579\, r^9 + 195\, r^8 + 585\, r^7 - 235\, r^6 + 431\, r^5 - 213\, r^4 + 101\, r^3 - 39\, r^2 + 16\, r - 10}{2\, \left(r^2 + 2\right)\, \left(r^{15} + 5\, r^{14} + 2\, r^{13} + 12\, r^{12} + 47\, r^{11} - 129\, r^{10} - 88\, r^9 - 122\, r^8 + 15\, r^7 + 99\, r^6 + 26\, r^5 + 96\, r^4 + r^3 + 25\, r^2 - 4\, r + 14\right)}\\ \frac{ - r^{16} + 3\, r^{15} + 58\, r^{14} + 131\, r^{13} + 214\, r^{12} + 487\, r^{11} + 122\, r^{10} + 551\, r^9 - 156\, r^8 + 489\, r^7 - 178\, r^6 + 361\, r^5 - 54\, r^4 + 45\, r^3 - 2\, r^2 - 19\, r - 3}{2\, \left(r^2 + 2\right)\, \left(r^{15} + 5\, r^{14} + 2\, r^{13} + 12\, r^{12} + 47\, r^{11} - 129\, r^{10} - 88\, r^9 - 122\, r^8 + 15\, r^7 + 99\, r^6 + 26\, r^5 + 96\, r^4 + r^3 + 25\, r^2 - 4\, r + 14\right)}\\ \frac{\left(r + 1\right)\, \left(r^{15} + 3\, r^{14} + 15\, r^{13} + 52\, r^{12} + 117\, r^{11} + 49\, r^{10} + 467\, r^9 - 278\, r^8 + 675\, r^7 - 427\, r^6 + 573\, r^5 - 336\, r^4 + 167\, r^3 - 73\, r^2 + 33\, r - 14\right)}{4\, \left(r^2 + 2\right)\, \left(r - 1\right)\, \left(r^{13} + 5\, r^{12} + 3\, r^{11} + 17\, r^{10} + 50\, r^9 - 112\, r^8 - 38\, r^7 - 234\, r^6 - 23\, r^5 - 135\, r^4 + 3\, r^3 - 39\, r^2 + 4\, r - 14\right)}\\ -\frac{2\, \left(r^2 + 1\right)\, \left(2\, r^{12} + 11\, r^{11} + 14\, r^{10} + 31\, r^9 + 68\, r^8 + 41\, r^7 + 73\, r^6 - 23\, r^5 + 31\, r^4 + 5\, r^2 + 4\, r - 1\right)}{\left(r^2 + 2\right)\, \left(r + 1\right)\, \left(r^{13} + 5\, r^{12} + 3\, r^{11} + 17\, r^{10} + 50\, r^9 - 112\, r^8 - 38\, r^7 - 234\, r^6 - 23\, r^5 - 135\, r^4 + 3\, r^3 - 39\, r^2 + 4\, r - 14\right)}\\ \frac{{\left(r - 1\right)}^2\, \left( - r^{13} + 17\, r^{11} + 55\, r^{10} + 84\, r^9 + 89\, r^8 + 106\, r^7 + 146\, r^6 + 63\, r^5 + 142\, r^4 - 27\, r^3 + 55\, r^2 + 14\, r + 25\right)}{2\, \left(r^2 + 2\right)\, \left( - r^{14} - 6\, r^{13} - 8\, r^{12} - 20\, r^{11} - 67\, r^{10} + 62\, r^9 + 150\, r^8 + 272\, r^7 + 257\, r^6 + 158\, r^5 + 132\, r^4 + 36\, r^3 + 35\, r^2 + 10\, r + 14\right)}\\ \frac{r^{16} + 6\, r^{15} + 28\, r^{14} + 45\, r^{13} + 101\, r^{12} - 108\, r^{11} + 826\, r^{10} - 317\, r^9 + 1397\, r^8 - 666\, r^7 + 1032\, r^6 - 685\, r^5 + 531\, r^4 - 256\, r^3 + 162\, r^2 - 67\, r + 18}{4\, \left(r^2 + 2\right)\, \left(r^{14} + 4\, r^{13} - 2\, r^{12} + 14\, r^{11} + 33\, r^{10} - 162\, r^9 + 74\, r^8 - 196\, r^7 + 211\, r^6 - 112\, r^5 + 138\, r^4 - 42\, r^3 + 43\, r^2 - 18\, r + 14\right)}\\ \frac{2\, r^{15} - 5\, r^{14} + 9\, r^{13} - 84\, r^{12} + 2\, r^{11} - 429\, r^{10} - 23\, r^9 - 444\, r^8 + 94\, r^7 - 143\, r^6 - 53\, r^5 + 76\, r^4 - 34\, r^3 + r^2 + 3\, r + 4}{2\, \left(r^2 + 2\right)\, \left(r^{14} + 4\, r^{13} - 2\, r^{12} + 14\, r^{11} + 33\, r^{10} - 162\, r^9 + 74\, r^8 - 196\, r^7 + 211\, r^6 - 112\, r^5 + 138\, r^4 - 42\, r^3 + 43\, r^2 - 18\, r + 14\right)}\\ \frac{r^{16} + 10\, r^{15} + 35\, r^{14} + 78\, r^{13} + 181\, r^{12} + 200\, r^{11} + 191\, r^{10} - 8\, r^9 + 135\, r^8 - 110\, r^7 + 241\, r^6 - 122\, r^5 + 183\, r^4 - 36\, r^3 + 45\, r^2 - 12\, r + 12}{2\, \left(r^2 + 2\right)\, \left( - r^{14} - 6\, r^{13} - 8\, r^{12} - 20\, r^{11} - 67\, r^{10} + 62\, r^9 + 150\, r^8 + 272\, r^7 + 257\, r^6 + 158\, r^5 + 132\, r^4 + 36\, r^3 + 35\, r^2 + 10\, r + 14\right)}\\ -\frac{ - r^{15} + 6\, r^{14} + 31\, r^{13} + 100\, r^{12} + 102\, r^{11} + 291\, r^{10} + 94\, r^9 + 305\, r^8 + 35\, r^7 + 156\, r^6 - 149\, r^5 + 142\, r^4 - 104\, r^3 + 27\, r^2 - 8\, r - 3}{2\, \left(r^2 + 2\right)\, \left(r^{14} + 4\, r^{13} - 2\, r^{12} + 14\, r^{11} + 33\, r^{10} - 162\, r^9 + 74\, r^8 - 196\, r^7 + 211\, r^6 - 112\, r^5 + 138\, r^4 - 42\, r^3 + 43\, r^2 - 18\, r + 14\right)}\\ -\frac{\left(r^2 + 1\right)\, \left(r^{14} + 9\, r^{13} + 29\, r^{12} + 67\, r^{11} + 140\, r^{10} + 86\, r^9 + 180\, r^8 - 6\, r^7 + 77\, r^6 - 119\, r^5 + 65\, r^4 - 29\, r^3 + 22\, r^2 - 8\, r - 2\right)}{\left(r^2 + 2\right)\, \left(r - 1\right)\, \left(r + 1\right)\, \left(r^{13} + 5\, r^{12} + 3\, r^{11} + 17\, r^{10} + 50\, r^9 - 112\, r^8 - 38\, r^7 - 234\, r^6 - 23\, r^5 - 135\, r^4 + 3\, r^3 - 39\, r^2 + 4\, r - 14\right)}\\ \frac{r^{16} - 4\, r^{15} - 20\, r^{14} - 55\, r^{13} + 89\, r^{12} + 574\, r^{11} + 1102\, r^{10} + 1467\, r^9 + 1029\, r^8 + 1248\, r^7 + 144\, r^6 + 607\, r^5 - 201\, r^4 + 230\, r^3 - 74\, r^2 + 29\, r - 22}{4\, r\, \left(r^2 + 2\right)\, \left( - r^{14} - 6\, r^{13} - 8\, r^{12} - 20\, r^{11} - 67\, r^{10} + 62\, r^9 + 150\, r^8 + 272\, r^7 + 257\, r^6 + 158\, r^5 + 132\, r^4 + 36\, r^3 + 35\, r^2 + 10\, r + 14\right)} \end{array}\right) \nonumber
\end{eqnarray}
corresponding to the product extrema given by the columns of the following matrix:
\begin{eqnarray}
\left(\begin{array}{ccccccccccccccc} 1 & 1 & 1 & 1 & 1 & 1 & 1 & 1 & 1 & 1 & 1 & 1 & 1 & 1 & 1\\ 1 & -1 & -1 & -1 & 1 & - r & 1 & - r & r & 1 & 1 & 1 & 1 & -1 & 1\\ r & 1 & 1 & - r & 1 & 1 & -1 & 1 & -1 & 1 & - r & -1 & 1 & r & - r\\ 1 & - r & - r & 1 & - r & 1 & r & 1 & 1 & r & -1 & - r & r & 1 & 1\\ 1 & -1 & -1 & 1 & r & 1 & 1 & - r & 1 & 1 & 1 & r & 1 & r & 1\\ 1 & 1 & 1 & -1 & r & - r & 1 & r^2 & r & 1 & 1 & r & 1 & - r & 1\\ r & -1 & -1 & - r & r & 1 & -1 & - r & -1 & 1 & - r & - r & 1 & r^2 & - r\\ 1 & r & r & 1 & - r^2 & 1 & r & - r & 1 & r & -1 & - r^2 & r & r & 1\\ r & 1 & 1 & 1 & -1 & 1 & - r & 1 & 1 & r & -1 & -1 & -1 & 1 & r\\ r & -1 & -1 & -1 & -1 & - r & - r & - r & r & r & -1 & -1 & -1 & -1 & r\\ r^2 & 1 & 1 & - r & -1 & 1 & r & 1 & -1 & r & r & 1 & -1 & r & - r^2\\ r & - r & - r & 1 & r & 1 & - r^2 & 1 & 1 & r^2 & 1 & r & - r & 1 & r\\ 1 & - r & r & - r & 1 & r & -1 & 1 & - r & 1 & - r & -1 & r & -1 & 1\\ 1 & r & - r & r & 1 & - r^2 & -1 & - r & - r^2 & 1 & - r & -1 & r & 1 & 1\\ r & - r & r & r^2 & 1 & r & 1 & 1 & r & 1 & r^2 & 1 & r & - r & - r\\ 1 & r^2 & - r^2 & - r & - r & r & - r & 1 & - r & r & r & r & r^2 & -1 & 1 \end{array}\right) \nonumber
\end{eqnarray}

\section{Case 3}

For Case 3, in the vicinity of $r=1/2$, and for a noise level given by
\begin{equation}
\lambda = 1 - {1 \over 1 + 2r}
\end{equation}
the output state of ${\cal C}_\lambda$ is given by (using the column vector notation of equation (\ref{column})):
\begin{eqnarray}
1/(1+2r) \left(\begin{array}{c} 1+2r \\ r \\ 1 \\ 1 \\ r \\ 1 \\ - r \\ r \\ 1 \\ - r \\ r^2 \\ 1 \\ 1 \\ r \\ 1 \\ 1 \end{array}\right) \nonumber
\end{eqnarray}
The probabilities $p_m$ appearing in the separable decomposition are given by:
 \begin{eqnarray}
\left(\begin{array}{c} \frac{{\left(r - 1\right)}^2\, \left( - r^6 + r^5 + 10\, r^3 + 7\, r^2 + 13\, r - 6\right)}{4\, \left(2\, r + 1\right)\, {\left(r + 1\right)}^2\, \left( - r^5 + r^4 + 2\, r^2 + r + 1\right)}\\ -\frac{ - 2\, r^9 + r^8 + 8\, r^7 - 4\, r^6 + 12\, r^5 - 18\, r^4 + 32\, r^3 - 60\, r^2 + 46\, r - 15}{4\, \left(2\, r + 1\right)\, {\left(r + 1\right)}^3\, \left( - r^5 + r^4 + 2\, r^2 + r + 1\right)}\\ 0\\ \frac{{\left(r - 1\right)}^2\, \left( - r^4 + 4\, r^2 + 8\, r + 1\right)}{4\, \left(2\, r + 1\right)\, \left( - r^5 + r^4 + 2\, r^2 + r + 1\right)}\\ \frac{{\left(r - 1\right)}^2\, \left( - 3\, r^6 + r^4 + 8\, r^3 + r^2 + 1\right)}{2\, r\, \left(2\, r + 1\right)\, {\left(r + 1\right)}^2\, \left( - r^5 + r^4 + 2\, r^2 + r + 1\right)}\\ -\frac{\left(r - 1\right)\, \left( - 2\, r^5 + r^4 + 4\, r^3 + 10\, r^2 + 2\, r + 1\right)}{\left(2\, r + 1\right)\, \left(r^2 + 1\right)\, {\left(r + 1\right)}^2\, \left( - r^3 + r^2 + r + 1\right)}\\ -\frac{ - r^7 + r^5 + 10\, r^4 + r^3 - 4\, r^2 - 9\, r + 2}{4\, r\, \left(2\, r + 1\right)\, {\left(r + 1\right)}^2\, \left( - r^3 + r^2 + r + 1\right)}\\ \frac{\frac{r^8}{2} - 3\, r^7 + 2\, r^6 + 6\, r^4 - 7\, r^3 + 2\, r - \frac{1}{2}}{r\, \left(2\, r + 1\right)\, {\left(r + 1\right)}^2\, \left( - r^5 + r^4 + 2\, r^2 + r + 1\right)}\\ \frac{\frac{r}{4} + \frac{\frac{r^2}{2} + r - \frac{3}{2}}{ - r^4 + 2\, r^2 + 2\, r + 1} + \frac{1}{4}}{2\, r + 1}\\ \frac{r\, \left( - r^5 - r^4 + 2\, r^3 + 6\, r^2 + 7\, r - 5\right)}{2\, \left(2\, r + 1\right)\, \left(r^2 + 1\right)\, \left(r + 1\right)\, \left( - r^3 + r^2 + r + 1\right)}\\ -\frac{2\, r^8 - 5\, r^7 + 7\, r^6 - 9\, r^5 + 7\, r^4 - 19\, r^3 + 9\, r^2 + r - 1}{4\, r^2\, \left(2\, r + 1\right)\, \left( - r^5 + r^4 + 2\, r^2 + r + 1\right)}\\ \frac{ - 2\, r^9 + 3\, r^8 - 8\, r^7 + 16\, r^6 - 8\, r^5 + 30\, r^4 - 24\, r^3 + 8\, r^2 + 2\, r - 1}{4\, r^2\, \left(2\, r + 1\right)\, \left(r^2 + 1\right)\, \left(r + 1\right)\, \left( - r^3 + r^2 + r + 1\right)}\\ \frac{ - 2\, r^{10} + 2\, r^9 - 3\, r^8 + 6\, r^7 + 6\, r^6 + 14\, r^5 + 20\, r^4 - 30\, r^3 + 28\, r^2 - 8\, r - 1}{4\, r\, \left(2\, r + 1\right)\, {\left(r + 1\right)}^2\, \left( - r^5 + r^4 + 2\, r^2 + r + 1\right)}\\ -\frac{ - 2\, r^{11} + 2\, r^{10} + 3\, r^9 + 3\, r^8 + 16\, r^7 - 16\, r^6 + 14\, r^5 - 58\, r^4 + 58\, r^3 - 26\, r^2 + 7\, r - 1}{4\, r\, \left(2\, r + 1\right)\, {\left(r + 1\right)}^3\, \left( - r^5 + r^4 + 2\, r^2 + r + 1\right)}\\ \frac{1}{r + 1} - \frac{7}{4\, \left(2\, r + 1\right)} + \frac{1}{4} \end{array}\right)
\end{eqnarray}
corresponding to the product extrema given by the columns of the following matrix:
\begin{eqnarray}
 = \left(\begin{array}{ccccccccccccccc} 1 & 1 & 1 & 1 & 1 & 1 & 1 & 1 & 1 & 1 & 1 & 1 & 1 & 1 & 1\\ 1 & 1 & -1 & -1 & -1 & - r & 1 & -1 & -1 & -1 & 1 & 1 & 1 & 1 & 1\\ 1 & 1 & 1 & 1 & -1 & 1 & - r & r & - r & 1 & 1 & 1 & 1 & 1 & -1\\ - r & - r & r & - r & r & 1 & 1 & -1 & 1 & r & r & r & - r & - r & - r\\ 1 & -1 & 1 & -1 & 1 & - r & 1 & 1 & 1 & - r & 1 & 1 & -1 & -1 & 1\\ 1 & -1 & -1 & 1 & -1 & r^2 & 1 & -1 & -1 & r & 1 & 1 & -1 & -1 & 1\\ 1 & -1 & 1 & -1 & -1 & - r & - r & r & - r & - r & 1 & 1 & -1 & -1 & -1\\ - r & r & r & r & r & - r & 1 & -1 & 1 & - r^2 & r & r & r & r & - r\\ -1 & -1 & -1 & 1 & 1 & 1 & 1 & 1 & 1 & 1 & r & - r & r & r & - r\\ -1 & -1 & 1 & -1 & -1 & - r & 1 & -1 & -1 & -1 & r & - r & r & r & - r\\ -1 & -1 & -1 & 1 & -1 & 1 & - r & r & - r & 1 & r & - r & r & r & r\\ r & r & - r & - r & r & 1 & 1 & -1 & 1 & r & r^2 & - r^2 & - r^2 & - r^2 & r^2\\ - r & r & r & - r & - r & 1 & r & - r & r & -1 & 1 & 1 & 1 & -1 & -1\\ - r & r & - r & r & r & - r & r & r & - r & 1 & 1 & 1 & 1 & -1 & -1\\ - r & r & r & - r & r & 1 & - r^2 & - r^2 & - r^2 & -1 & 1 & 1 & 1 & -1 & 1\\ r^2 & - r^2 & r^2 & r^2 & - r^2 & 1 & r & r & r & - r & r & r & - r & r & r \end{array}\right) \nonumber
 \end{eqnarray}

\section{Case 4}

The separable decomposition in Case 4 was guessed by hand, and hence is simpler that the expressions derived above using numerical and computational
algebra techniques. For Case 4 and a noise level given by:
\begin{equation}
\lambda = 1 - {1 \over 2 +r^2} \nonumber
\end{equation}
the output state of ${\cal C}_\lambda$ is given by (using the column vector notation of equation (\ref{matrix})):
\begin{eqnarray}{1 \over 2 +r^2} \left(\begin{array}{cccc}
      2 +r^2 & r & 1 & 1 \\
     1 & r & -r^2 & 1  \\
     r & -1 & r & r  \\
     1 & r & 1 & 1
      \end{array}\right)
\end{eqnarray}
The separable decomposition of this state is given by the following expression, which works for all values of $r$ (as the probabilities
are always positive):
\begin{eqnarray}
{1 \over 2 +r^2}\left(\begin{array}{cccc}
      1 & r & 1 & 1 \\
     1 & r & 1 & 1  \\
     r & r^2 & r & r  \\
     1 & r & 1 & 1
      \end{array}\right)
+
{1 +r^2 \over 2 +r^2} \left(\begin{array}{cccc}
      1 & 0 & 0 & 0 \\
     0 & 0 & -1 & 0  \\
     0 & -1 & 0 & 0 \\
     0 & 0 & 0 & 0
      \end{array}\right)
\end{eqnarray}
where the rightmost matrix is simply a uniform mixture of all product vectors $(x_A,y_A,z_A)\otimes(x_B,y_B,z_B)$ that satisfy
$|x_A|=|x_B|=|y_A|=|y_B|=1$, $|z_A|=|z_B|=r$, and the anticorrelation condition $x_A=-y_B$ and $x_B=-y_A$ (which ensures the presence
of the central $-1$ elements).

\end{widetext}

\end{document}